\documentclass[english,preprint,superscriptaddress,amsmath,amssymb,a4paper]{revtex4}
\usepackage[T1]{fontenc}
\usepackage[latin1]{inputenc}
\usepackage{graphicx}
\usepackage[top=0.95in,bottom=0.95in,,left=0.85in,right=0.85in] {geometry}
\usepackage{booktabs}
\usepackage{color}
\usepackage{jabbrv}
\makeatletter
\usepackage{dcolumn}
\usepackage{bm}
\usepackage{epsfig}

\newcommand{\ignore}[1]{}

\usepackage{babel}
\makeatother

\usepackage{sidecap}
\sidecaptionvpos{figure}{c}

\begin{document}

\title{Higher-Order Topology in Monolayer FeSe}

\author{Gan Zhao}
\affiliation{Hefei National Laboratory for Physical Sciences at the Microscale,
  CAS Key Laboratory of Strongly-Coupled Quantum Matter Physics, Department of Physics,
  University of Science and Technology of China, Hefei, Anhui 230026, China}

\author{Haimen Mu}
\affiliation{Hefei National Laboratory for Physical Sciences at the Microscale,
  CAS Key Laboratory of Strongly-Coupled Quantum Matter Physics, Department of Physics,
  University of Science and Technology of China, Hefei, Anhui 230026, China}

\author{Huimin Zhang}
\affiliation{Department of Physics and Astronomy,
  West Virginia University, Morgantown, WV, 26506, USA}

\author{Z. F. Wang}\thanks{E-mail: zfwang15@ustc.edu.cn}
\affiliation{Hefei National Laboratory for Physical Sciences at the Microscale,
  CAS Key Laboratory of Strongly-Coupled Quantum Matter Physics, Department of Physics,
  University of Science and Technology of China, Hefei, Anhui 230026, China}




\begin{abstract}
Generally, the topological corner state in two-dimensional second-order topological insulator (2D SOTI) is equivalent
to the well-known domain wall state, originated from the mass-inversion between two adjacent edges with phase shift of $\pi$.
In this work, go beyond this conventional physical picture, we report a fractional mass-kink induced 2D SOTI in monolayer FeSe
with canted checkerboard antiferromagnetic (AFM) order by analytic model and first-principles calculations. The canted spin
associated in-plane Zeeman field can gap out the quantum spin Hall edge state of FeSe, forming a fractional mass-kink
with phase shift of $\pi/2$ at the rectangular corner, and generating an in-gap topological corner state with fractional charge of e/4.
Moreover, the topological corner state is robust to local perturbation, existing in both naturally and non-naturally cleaved corners,
regardless of the edge orientation. Our results not only demonstrate a material system to realize the unique 2D AFM SOTI,
but also pave a new way to design the higher-order topological states from fractional mass-kink with arbitrary phase shift,
which are expected to draw immediate experimental attention.\\

\end{abstract}

\maketitle

\clearpage
\noindent \textbf{INTRODUCTION}\\
With the rapid progress in classification of topological electronic states \cite{1,2}, different kinds of topological materials
are predicted theoretically and confirmed experimentally \cite{3,4,5}. However, the prior studies are mainly focused on the first-order
topological materials \cite{6,7,8}, where the topological boundary state only appears at dimension one lower than that of
the bulk state. When this conventional bulk boundary correspondence is extended to the higher-order form \cite{9}, a new class of topological
materials called higher-order topological insulator emerges \cite{10,11,12,13}. The \textit{m}-dimensional \textit{n}th-order topological
insulator holds gapless state at (\textit{m-n})-dimensional boundary but gapped state otherwise. Therefore, the two-dimensional
second-order topological insulator (2D SOTI) is characterized by gapped topological edge state and in-gap topological corner state.
Different to symmetry protected bulk topological index, physically, the emergence of in-gap topological corner state can also be
understood by the well-known mass-inversion mechanism in Jackiw-Rebbi model \cite{14}, which can support a localized domain wall state,
as shown schematically in Fig. 1(a).

\begin{figure}
  \centering
  \includegraphics[width=13cm]{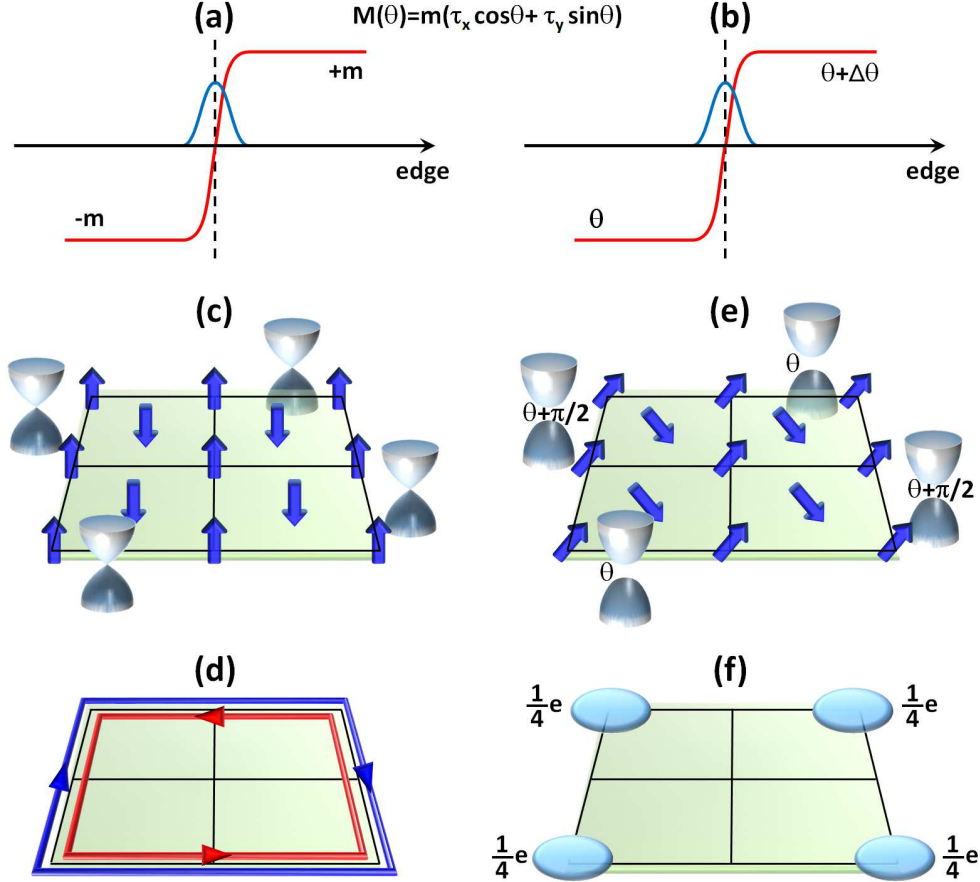}
  \caption{Schematic mass distribution along the edge (red line) and localized domain wall state (blue line) for (a) mass-inversion,
    (b) fractional mass-kink. The mass \textit{M} is defined in edge state space, \textit{m} is mass-intensity,
    $\theta$ is mass-angle, $\Delta\theta$ is phase shift across the domain wall, $\tau_x$ and $\tau_y$ are Pauli matrix of the edge state.
    (c) Schematic checkerboard AFM order in monolayer FeSe and gapless topological edge state. (d) Schematic real-space distribution of gapless
    topological edge state (QSH edge state), showing the propagating direction (arrow) and spin (red and blue color).
    (e) Schematic canted checkerboard AFM order in monolayer FeSe and gapped topological edge state. The adjacent edges have a different
    mass-angle of $\theta$ and $\theta+\pi/2$. (f) Schematic real-space distribution of in-gap topological corner state,
    holding a fractional charge of e/4.}
\end{figure}

Currently, although a variety of theoretical models are proposed for the 2D SOTI \cite{15,16,17,18,19,20,21,22,23,24,25,26,27,28,29},
its material realization is still a challenging task. Besides some artificial macro-structures \cite{30,31,32,33,34,35,36},
few realistic electronic materials have been reported. What's more, the limited 2D SOTI materials, such as graphdiyne \cite{39,40},
$\gamma$-graphyne \cite{41} and twisted-bilayer graphene \cite{42,43,45}, are all describable by the mass-inversion mechanism.
Recently, this physical mechanism is extended to a more universal form for generating the domain wall state \cite{46},
as shown schematically in Fig. 1(b). The mass distribution along the edge is characterized by the mass-angle, which is $\theta$ and
$\theta+\Delta\theta$ for left- and right-part, respectively. Based on model calculations, the localized domain wall state is achieved
in this fractional mass-kink \cite{47} with any non-zero phase shift ($\Delta\theta \neq$ 0). Obviously, the mass-inversion in Fig. 1(a)
just corresponds to one special case of $\Delta\theta=\pi$ in Fig. 1(b). The fractional mass-kink further lowers the physical
requirement to realize the domain wall state, providing much freedom to design the 2D SOTI. However, no material systems
are reported for this peculiar non-$\pi$ phase shift associated topological corner state, greatly hindering the experimental verification.

Due to the tunable topological phase transition between first- and higher-order topological states \cite{48,49}, the quantum spin
Hall (QSH) materials provide a good platform to explore the 2D SOTI. As the parent compound of Fe-based superconductor,
the monolayer FeSe has been intensively studied \cite{50}. Recently, the QSH state with checkerboard antiferromagnetic (AFM) order (Fig. 1(c))
is confirmed experimentally in monolayer FeSe \cite{51}, demonstrating an exotic topological phase in this high-temperature superconductor \cite{52,53}.
In this work, starting from the AFM QSH state, we report a 2D SOTI in monolayer FeSe with canted checkerboard AFM order (Fig. 1(e)).
An intriguing fractional mass-kink induced topological corner state is identified by analytic model and first-principles calculations,
which is robust to local perturbation and edge orientation. The main discoveries of our work are summarized
in Fig. 1. Since the AFM QSH state has a pair of helical Dirac edge states (Fig. 1(c) and 1(d)), the coupling between them
will break the band degeneracy at the Dirac point. Physically, this can be achieved by the canted spin induced
in-plane Zeeman field. Meanwhile, the fractional mass-kink with phase shift of $\pi/2$ is formed at the rectangular corner (Fig. 1(e)),
resulting in an in-gap topological corner state with fractional charge of e/4 (Fig. 1(f)).\\

\noindent \textbf{RESULTS}\\
The nontrivial band topology of monolayer FeSe with checkerboard AFM order is captured by its band structures around M point
in the first Brillouin zone \cite{51}. The low-energy effective Hamiltonian derived from the first-principle
calculations is written as \cite{54}
\begin{equation}
  \begin{split}
    H&=a_0(k_x^2+k_y^2)s_0\sigma_0+a_1k_xk_y s_0\sigma_x \\
    & \quad +\lambda s_z\sigma_y +a_2(k_x^2-k_y^2)s_0\sigma_z
  \end{split}
\end{equation}
where $\sigma_0$ and $s_0$ are identity matrix, $\sigma_{x,y,z}$ is orbital Pauli matrix, $s_z$ is spin Pauli matrix,
$a_{0,1,2}$ is fitting constant, and $\lambda$ is intrinsic spin-orbital coupling (SOC). The characterized band structures
of Eq. (1) are shown in Fig. 2, exhibiting a fourfold rotational symmetry ($C_4= i\sigma_y e^{i\pi s_z/4}$). Without SOC,
there is a quadratic band degeneracy at M point (Fig. 2(a)). With SOC, the degeneracy is lift and an AFM QSH
state is realized (Fig. 2(b)). One can see that this effective Hamiltonian is dramatically different to Kane-Mele \cite{55}
and Bernevig-Hughes-Zhang \cite{56} model described QSH state that includes the linear term of momentum. The helical Dirac edge state
for such a quadratic QSH Hamiltonian \cite{57,58,59} has not been analytically derived yet.

\begin{figure}
  \centering
  \includegraphics[width=9cm]{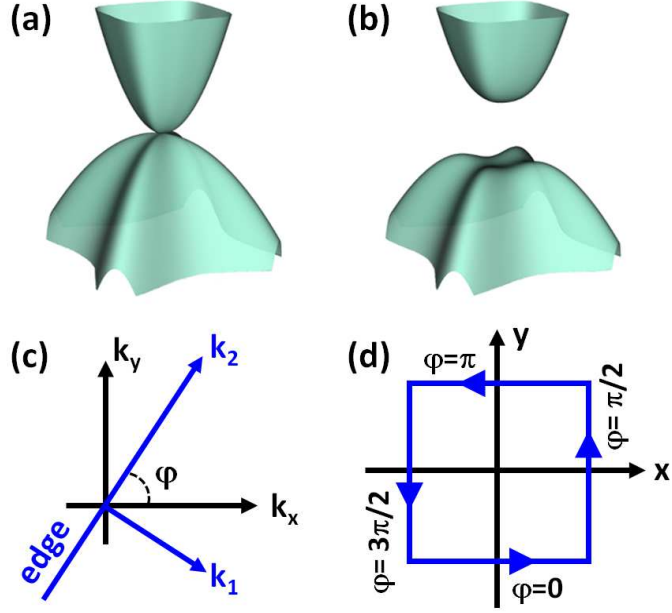}
  \caption{(a) and (b) Characterized band structures of the low-energy effective Hamiltonian without and with SOC, showing
    the quadratic touching point and quadratic QSH. (c) Reciprocal space rotated coordinate system for describing the edge along $k_2$
    direction that has a angle $\varphi$ with $k_x$ axis. (d) Real space rectangular cluster with four edges, corresponding to
    $\varphi=0,\pi/2,\pi$ and $3\pi/2$ in (c). The arrow denotes the positive direction of the edge.}
\end{figure}

The first term in Eq. (1) is a reference energy that doesn't affect our final results,
so we omit it in the following part. To simplify the derivation of the helical Dirac edge state along different edges,
we define a rotated coordinate system $k_1$-$k_2$, where the edge along $k_2$ direction has an angle $\varphi$
with $k_x$ axis, as shown in Fig. 2(c). For the rectangular cluster with $90^\circ$ corners, the four regular
edges are along the direction of $\varphi= 0, \pi/2, \pi$ and $3\pi/2$, as shown in Fig. 2(d). In the $k_1$-$k_2$ plane,
Eq. (1) can be rewritten as
\begin{equation}
  \begin{split}
    H_{\pm}=\mp a_1k_1k_2s_0\sigma_x+\lambda s_z\sigma_y \pm a_2(k_2^2-k_1^2)s_0\sigma_z \\
  \end{split}
\end{equation}
where $H_+$ for $\varphi$=0, $\pi$ and $H_-$ for $\varphi$=$\pi/2$, $3\pi/2$.
Assuming Eq. (2) is defined in half-space ($k_1$>0) of the $k_1$-$k_2$ plane, we replace $k_1 \rightarrow -i\partial_{x_1}$,
$k_2 \rightarrow 0$, and substitute a trial edge state solution $\psi=e^{\eta x_1}\phi$ into it, where $\eta$ is a decay
constant. Then, Eq. (2) is reduced to the form
\begin{equation}
  \begin{split}
    \widetilde{H}_{\pm}=\lambda s_z\sigma_y \pm a_2\eta^2s_0\sigma_z
  \end{split}
\end{equation}
The spin is decoupled in this bulk Hamiltonian, so the edge state solution will also be
spin-decoupled, which has a form of $|\psi_{\uparrow\pm}\rangle=(|\phi_{\uparrow\pm}\rangle,0)^T$ and $|\psi_{\downarrow\pm}\rangle=(0,|\phi_{\downarrow\pm}\rangle)^T$.
Substituting $|\psi_{\uparrow+}\rangle$ and $|\psi_{\downarrow+}\rangle$ into $\widetilde{H}_+$, considering the solution
divergence ($x_1\rightarrow +\infty$) and time-reversal symmetry ($T=is_yK$, $K$ is Hermitian conjugate),
we obtain the edge state solution of $\widetilde{H}_+$ as
\begin{equation}
  \begin{split}
    |\phi_{\uparrow+}\rangle&=\alpha e^{\eta x_1}|\xi_+\rangle + \beta e^{\eta^* x_1}|\xi_-\rangle \\
    |\phi_{\downarrow+}\rangle&=-|\phi_{\uparrow+}^*\rangle
  \end{split}
\end{equation}
where $\eta^2=i\lambda/a_2$, $\alpha$ and $\beta$ are coefficients, and $|\xi_\pm\rangle=\sqrt{2}/2(1,\pm1)^T$ are eigenstate of $\sigma_x$.
Since $\widetilde{H}_\pm$ are connected by the $C_4$ symmetry, that is $\widetilde{H}_- = C_4^{-1}\widetilde{H}_+C_4$, the edge state solution of
$\widetilde{H}_-$ can be obtained from the relation
\begin{equation}
  \begin{split}
    |\psi_{\uparrow-}\rangle &= C_4^{-1}|\psi_{\uparrow+}\rangle \\
    |\psi_{\downarrow-}\rangle &= C_4^{-1}|\psi_{\downarrow+}\rangle
  \end{split}
\end{equation}
In the edge state space ($|\psi_{\uparrow\pm}\rangle, |\psi_{\downarrow\pm}\rangle$),
to the leading order of $k_2$, Eq. (2) can be transformed into the standard Dirac equation as
\begin{equation}
  \begin{split}
    h_{\pm}=a_1k_2\rm{Im}(\eta)\tau_z
  \end{split}
\end{equation}
where $\tau_z$ is Pauli matrix of the edge state, showing a general form of
the helical Dirac edge state for the quadratic QSH Hamiltonian.

\begin{figure*}
  \includegraphics[width=16cm]{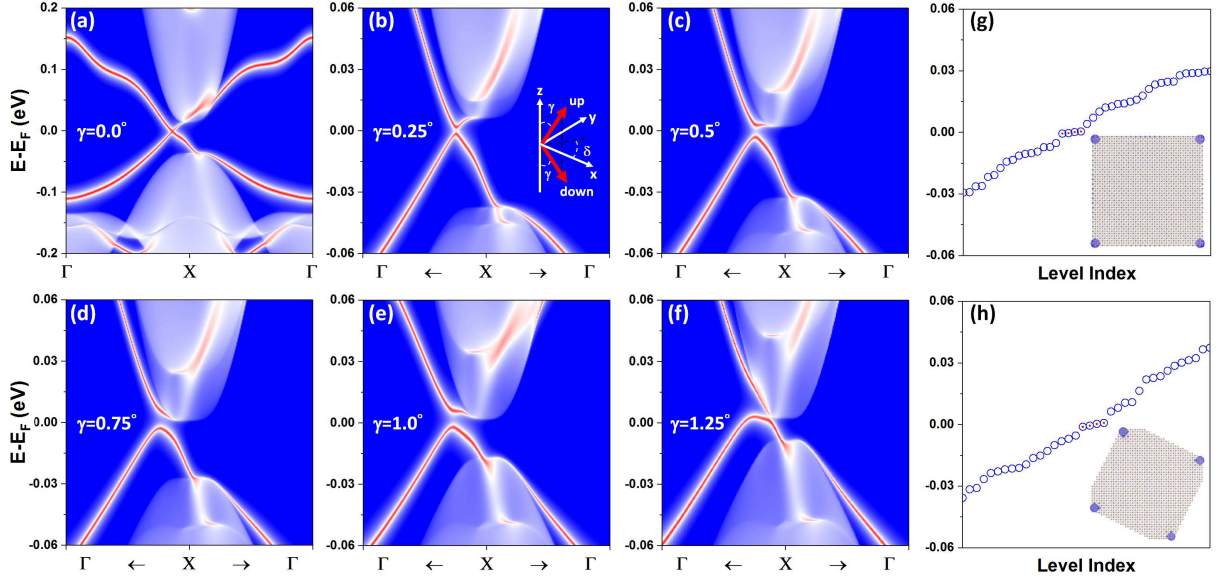}
  \caption{(a) The gapless QSH edge state along ferromagnetic edge of monolayer FeSe with checkerboard AFM order.
    (b-f) Zoom-in gapped topological edge state by canting the spin direction with $\delta=45^\circ$, (b) $\gamma=0.25^\circ$,
    (c) $\gamma=0.5^\circ$, (d) $\gamma=0.75^\circ$, (e) $\gamma=1.0^\circ$ and (f) $\gamma=1.25^\circ$. The canted spin-up
    and -down directions are determined by angle $\gamma$ and $\delta$, as shown schematically in inset of (b).
    (g) Discrete energy levels of rectangular monolayer FeSe cluster with four naturally cleaved ferromagnetic edges of the same spin.
    The four in-gap topological corner states are marked in red-dot around the Fermi-level. The inset is spatial charge density
    distribution of the corner states. The radii of circle on each atom denotes the absolute value of charge density.
    (h) is the same as (g), but for a rotated rectangular cluster with non-naturally cleaved ferromagnetic edges of the same spin.}
\end{figure*}

In order to gap out the helical Dirac edge state in Eq. (6), we consider an in-plane Zeeman field as \cite{48}
\begin{equation}
  \begin{split}
    H^\prime=us_x\sigma_0+vs_y\sigma_0
  \end{split}
\end{equation}
where $s_{x,y}$ is spin Pauli matrix, $u$ and $v$ are coefficients that control the intensity and direction of in-plane Zeeman field.
This new term breaks the spin-degeneracy and mixes the spin-up and -down in Eq. (1). Similar to the above derivation,
Eq. (7) can be written in the edge state space as \cite{46}
\begin{equation}
  \begin{split}
    h^\prime_+ &=m(\tau_x {\rm cos} \theta + \tau_y {\rm sin} \theta ) \\
    h^\prime_- &=m[\tau_x {\rm cos} (\theta+\pi/2) + \tau_y {\rm sin} (\theta+\pi/2) ]
  \end{split}
\end{equation}
where $h^\prime_+$ for $\varphi=0, \pi$ and $h^\prime_-$ for $\varphi=\pi/2, 3\pi/2$.
$\tau_{x,y}$ is Pauli matrix of the edge state, $m_1 = {\rm Re} (\frac{\alpha^2}{2 \eta} + \frac{\beta^2}{2 \eta^*})$,
$m_2 = {\rm Im} (\frac{\alpha^2}{2 \eta} + \frac{\beta^2}{2 \eta^*})$,
$m=[(u^2+v^2)(m_1^2+m_2^2)]^{1/2}$, $m {\rm{cos}} \theta=um_1-vm_2$ and $m {\rm{sin}} \theta=um_2+vm_1$.
Combining Eq. (6) and Eq. (8), the helical Dirac edge state is gapped out with a band gap of $E_g=2m$.
If the intensity of in-plane Zeeman field $(u^2+v^2)^{1/2}$ is fixed, the Dirac gap will independent of the direction of
in-plane Zeeman field and mass-angle. Most remarkably, one notices that the mass-angle is
$\theta$ and $\theta+\pi/2$ for edge along $\varphi=0$, $\pi$ and $\varphi=\pi/2$, $3\pi/2$, respectively, forming a
fractional mass-kink with phase shift of $\Delta\theta=\pi/2$ at each $90^\circ$ corner. According to Moore's theory \cite{46},
such a phase shift will support a topological corner state with fractional charge of $N_s=e|\Delta\theta/2\pi|=e/4$.
Therefore, a 2D SOTI is identified in the monolayer FeSe, originated from the in-plane Zeeman field induced fractional
mass-kink with non-$\pi$ phase shift. More details about the analytic derivations can be found in the Supplementary Data.

To further support our analytic results, the first-principles calculations are performed to directly identify
the unique gapped topological edge state and in-gap topological corner state in monolayer FeSe. The gapless QSH edge state along ferromagnetic
edge of [100] direction for monolayer FeSe with checkerboard AFM order is shown in Fig. 3(a), where the $PT$ symmetry
protected Dirac point is sitting along $\Gamma$-$\rm X$ direction \cite{51}. In order to introduce an in-plane Zeeman field to
gap out the topological edge state, a special canted checkerboard AFM order is considered (Fig. 1(e)) \cite{61}. The direction
of canted spin is determined by zenith angle ($\gamma$) and azimuth angle ($\delta$), as shown schematically in inset of Fig. 3(b).
Experimentally, this spin configuration can be realized by applying an in-plane magnetic field, making the spin
canted along the field direction, where zenith and azimuth angle is tunable by the field intensity and direction \cite{62}.
Interestingly, we found that the spectra of topological edge state with canted spin is insensitive to $\delta$ (see Fig. S1),
that is, it doesn't depend on the direction of in-plane Zeeman field, which is consistent with our analytic derivation. Without
losing the generality, we will fix $\delta$ and discuss the effect of $\gamma$ in the following part. For azimuth angle $\delta=45^\circ$
along [110] direction, the zoom-in topological edge state with small zenith angle ranging from $\gamma=0.25^\circ$ to $1.25^\circ$ is shown
in Fig. 3(b) to 3(f), respectively. Comparing to the gapless spectra shown in Fig. 3(a), there are two significant features can be
observed for the canted spin. First, the spin-degenerate bulk bands are split and bulk gap around X point is reduced with the
increasing of $\gamma$. This feature can be attributed to the in-plane Zeeman field induced spin-splitting that results
in upshift and downshift of the opposite spin bands (see Fig. S2). With the increasing of $\gamma$, the intensity of in-plane
Zeeman field is enhanced, so the band splitting is increased and bulk gap between opposite spins is reduced. Second, the topological
edge state is gapped out and Dirac gap exhibits a non-monotonic behavior with the increasing of $\gamma$. This feature can be
attributed to the in-plane Zeeman field induced mass-term and band-reshaping. According to our analytic results, the Dirac gap is
proportional to the intensity of in-plane Zeeman field. In principles, it will increase with the increasing of $\gamma$.
However, the reduced bulk gap moves the bottom branch of topological edge state gradually close to the top bulk band, making the
global Dirac gap decreased with further increasing of $\gamma$. Therefore, based on canted checkerboard AFM order,
we confirm the gapped topological edge state in monolayer FeSe, identifying the first unique character of 2D SOTI.

Since the band topology is same in the canted spin opened Dirac gap, we will focus on one gapped topological edge state with
$\gamma=0.75^\circ$ and $\delta=45^\circ$ (Fig. 3(d)) to illustrate its corner state (see Fig. S3(a)-(c)).
The rectangular cluster with $90^\circ$ corners is constructed by cutting four ferromagnetic edges with the same spin
along naturally cleaved [100] and [010] directions \cite{FeSe}. The discrete energy levels of the cluster are shown in Fig. 3(g).
Clearly, there are four nearly degenerate corner states around the Fermi-level, as labeled by the red-dot. The spatial distribution
of them is shown in the inset of Fig. 3(g), which is localized at four corners. Therefore, we confirm the in-gap topological edge
state in monolayer FeSe, identifying the second unique character of 2D SOTI. Furthermore, by accounting the total electrons in
the system \cite{39}, there is only one electron is left after filling all energy levels below the four corner states. Consequently,
each corner state will hold a fractional charge of e/4, which is consistent with our analytic derivation.

In order to facilitate the possible experimental measurement, the robustness of topological corner state against the
disorder and edge-cutting orientation are further investigated. The local disorder is simulated by adding finite random
on-site energy to atoms around the corner. There are still four corner states localized at the corners, but the relative
intensity is different between the corner with and without disorder (see Fig. S3(d)). Most importantly, we found
that the rectangular cluster constructed from non-naturally cleaved ferromagnetic edge along arbitrary orientations
can also support the existence of topological corner state. The discrete energy levels and spatial distribution of the corner states
are shown in Fig. 3(h) and its inset, which are comparable to those shown in Fig. 3(g) and its inset. Moreover, the same results are
obtained for the rectangular cluster with four antiferromagnetic edges (see Fig. S3(e)). Consequently, the topological corner
state in monolayer FeSe can be revealed by cutting the edge along irregular orientation, providing more convenience for scanning
tunneling microscopy detection. Additionally, if monolayer FeSe is stacked vertically into a three-dimensional (3D) structure of
(LiFe)OHFeSe \cite{63}, the angle-resolved photoemission spectroscopy measured band structures \cite{64} are similar to the monolayer
FeSe studied in this work \cite{51}. Considering the weak coupling between the neighboring FeSe layers in (LiFe)OHFeSe \cite{65},
a 3D weak SOTI \cite{66} is also realizable by the canted checkerboard AFM order, where the topological corner state is stacked into a one-dimensional
topological hinge state with little dispersion along the vertical direction, supporting a quantized hinge conductance (see Fig. S4).\\

\noindent \textbf{CONCLUSION}\\
In conclusion, based on analytic model and first-principles calculations, we identify an intriguing fractional
mass-kink induced topological corner state in monolayer FeSe with canted checkerboard AFM order.
Our results greatly extend the topological physics for mass-inversion induced domain wall state and
provide a new way to design the higher-order topological materials. The coupling between
topological corner state and superconducting state in monolayer FeSe will also offer an attractive
opportunity to realize the majorana fermion in the future.\\

\noindent \textbf{FUNDING}

\noindent This work was supported by NSFC (Grant No. 11774325 and 21603210), National Key Research and Development
Program of China (Grant No. 2017YFA0204904) and Fundamental Research Funds for the Central Universities.
We also thank Supercomputing Center at USTC for providing the computing resources.\\

\end{document}